\documentclass[twocolumn, trackchanges, twocolappendix]{aastex7}

\usepackage{natbib}
\usepackage{amsmath}
\usepackage{braket}
\usepackage{color}
\hypersetup{colorlinks=true, linkcolor=black, citecolor=blue,}

\def\rev#1{#1}

\begin{document}

\title{Empirical Optimization of the Source-Surface Height in the PFSS extrapolation}

\author[orcid=0000-0002-7136-8190, gname=Munehito, sname=Shoda]{Munehito Shoda} 
\affiliation{Department of Earth and Planetary Science, School of Science, The University of Tokyo, 7-3-1 Hongo, Bunkyo-ku, Tokyo, 113-0033, Japan}
\email{shoda.m.astroph@gmail.com}

\author[orcid=0009-0007-8101-9668, gname=Kyogo, sname=Tokoro]{Kyogo Tokoro} 
\affiliation{Department of Earth and Planetary Science, School of Science, The University of Tokyo, 7-3-1 Hongo, Bunkyo-ku, Tokyo, 113-0033, Japan}
\email{}

\author[orcid=0000-0002-9032-8792, gname=Daikou, sname=Shiota]{Daikou Shiota} 
\affiliation{National Institute of Information and Communications Technology, 4-2-1 Nukui-kita, Koganei,
Tokyo 184-8795, Japan}
\email{}

\author[orcid=0000-0001-7891-3916, gname=Shinsuke, sname=Imada]{Shinsuke Imada} 
\affiliation{Department of Earth and Planetary Science, School of Science, The University of Tokyo, 7-3-1 Hongo, Bunkyo-ku, Tokyo, 113-0033, Japan}
\email{}

\begin{abstract}
The potential field source surface (PFSS) method is a widely used magnetic field extrapolation technique in the space weather community. The only free parameter in the PFSS method is the source-surface height ($R_{\rm SS}$), beyond which all field lines are open. Although $R_{\rm SS}$ is known to vary with solar activity, there is no consensus on how to determine it for a given surface magnetic field distribution. In this study, we investigate the nature of $R_{\rm SS}$ using a long-period (2006–2023) data, covering two solar minima and one maximum. We adopt ADAPT-GONG magnetograms and determine $R_{\rm SS}$ by matching the open flux estimated from observations at 1 au with that calculated using the PFSS method. \rev{Our analysis reveals that $R_{\rm SS}$ increases slightly after the solar minima and around the solar maximum, and that it can be characterized by both the mean unsigned photospheric magnetic field strength and the dipolarity parameter $f_{\rm dip}$, defined as $f_{\rm dip} = B_{\rm dip}^2/(B_{\rm dip}^2 + B_{\rm quad}^2 + B_{\rm oct}^2)$, with $B_{\rm dip}$, $B_{\rm quad}$, and $B_{\rm oct}$ denoting the magnitudes of dipolar, quadrupolar, and octupolar components of photospheric radial magnetic field, respectively.} Our results suggest that $R_{\rm SS}$ does not exhibit a simple monotonic dependence on the solar activity and must be determined by properly considering both surface magnetic field strength and global field structure.
\end{abstract}

\section{Introduction} \label{sec:introduction}

Space weather forecasting is the prediction of the effects of solar magnetic activity on Earth and interplanetary space \citep{Cade_2015_SpWea, Temmer_2021_LRSP, Khabarova_2024_FrASS}, and its significance has been increasingly acknowledged in recent years \citep{Lanzerotti_2017_SSRev, Xue_2024_SpWea}. The solar wind is the primary constituent of interplanetary space, influencing the propagation of coronal mass ejections \citep{Gopalswamy_2000_GeoRL, Temmer_2011_ApJ, Wu_2024_ApJ} and the formation of co-rotating interaction regions \citep{Gosling_1999_SSRv, Richardson_2018_LRSP}, both of which impact geomagnetic activity. Thus, understanding its spatiotemporal structure is vital for space weather forecasting.

\begin{deluxetable*}{lcccc}
\tablewidth{0pt}
\tablecaption{List of previous studies on source-surface height optimization \label{table:previous_works}}
\tablehead{
\colhead{Reference} & \colhead{Epoch} & \colhead{Photospheric field} & \colhead{Metric} & \colhead{Optimal $R_{\rm SS}/R_\odot$}
}
\startdata
\citet{Hoeksema_1983_JGR} & 1976-1982 & WSO & \begin{tabular}{c} IMF polarity \\ (observed at 1 au) \end{tabular} & 2.5 \\ \hline
\citet{Lee_2011_SoPh} & \begin{tabular}{l} 1994-1996 \\ 2006-2009 \end{tabular} & MWO & \begin{tabular}{c} open flux \\ (observed at 1 au) \end{tabular} & 1.8-1.9 \\ \hline
\citet{Arden_2014_JGR} & 1996-2011 & \begin{tabular}{c} flux-transport model \\ (LMSAL) \end{tabular} & \begin{tabular}{c} open flux \\ (observed at 1 au) \end{tabular}  & \begin{tabular}{c} 2.2-3.3 \\ (larger in solar min.) \end{tabular} \\ \hline
\citet{Nikolic_2019_SpWea} & 2006-2018 & GONG & \begin{tabular}{c} open flux \\ (observed at 1 au) \end{tabular}  & \begin{tabular}{c} $\lesssim 1.5$ \\ (larger in solar max.) \end{tabular} \\ \hline
\citet{Badman_2022_ApJ} & 2019-2020 &  \begin{tabular}{c} ADAPT-GONG \\ ADAPT-HMI \end{tabular} & coronal hole area & $\sim$ 2.0 \\ \hline
\citet{Badman_2022_ApJ} & 2019-2020 &  \begin{tabular}{c} ADAPT-GONG \\ ADAPT-HMI \end{tabular} & \begin{tabular}{c} location of \\ streamer belt \end{tabular} & $\geq$ 3.0 \\ \hline
\citet{Wagner_2022_AandA} & \begin{tabular}{c} 2008 Aug. \\ 2010 Jul. \end{tabular} & ADAPT-GONG & \begin{tabular}{c} feature matching \\ $\&$ coronal hole area \end{tabular} &  $\sim$ 2.0$^{(1)}$  \\ \hline 
\citet{Huang_2024_ApJ} & 2011-2019 & ADAPT-GONG & \begin{tabular}{c} open-field area \\ (AWSoM model) \end{tabular}  & \begin{tabular}{c} 1.6-2.7 \\ (larger in solar max.) \end{tabular} \\ \hline
\citet{Benavitz_2024_ApJ} & 2008-2020 & ADAPT-GONG & field-line angle  & \begin{tabular}{c} 1.3-3.0 \\ (larger in solar min.) \end{tabular} \\ \hline
\enddata
\tablecomments{(1) We adopt the value of the inner boundary of the Scatten current sheet (SCS) model, denoted $R_{\rm SCS}$ in the original paper. This is because, in PFSS–SCS combined extrapolation, the open flux is governed by $R_{\rm SCS}$—the interface between PFSS and SCS—rather than by the outer boundary $R_{\rm SS}$.}
\end{deluxetable*}

The solar wind originates in the corona \citep{Parker_1958_ApJ, Velli_1994_ApJ}, where magnetic field governs the energetics \citep{Gary_2001_SolPhys, Iwai_2014_EPS}, making its properties strongly dependent on the magnetic field geometry \citep{Wang_1990_ApJ, Riley_2001_JGR, Arge_2004_JASTP, Tokumaru_2024_SoPh}. Therefore, knowing the coronal magnetic field is essential for space weather forecasting \citep{Odstrcil_2003_AdSpR, Shiota_2014_SpaceWeather, Pomoell_2018_JSWSC} and identifying the solar wind source region \citep{Neugebauer_1998_JGR, Stansby_2021_SoPh, Koukras_2025_AandA}. The coronal magnetic field structure is also essential in the flux-tube models for studying solar wind acceleration \citep{Kopp_1976_SolPhys, Shoda_2018_ApJ_a_self-consistent_model, Chandran_2019_JPP, Shoda_2022_ApJ}.

Since direct observations of the coronal magnetic field remain limited \citep{Lin_2004_ApJ, Jess_2016_NatPhys, Schad_2024_SciA}, its global structure \rev{can} be inferred by extrapolating from the photosphere. Several extrapolation methods have been proposed\rev{, including potential field source surface (PFSS) model \citep{Altschuler_1969_SolPhys, Schatten_1969_SolPhys}, Schatten current sheet (SCS) model \citep{Schatten_1971, Schatten_1972_NASSP}, current sheet source surface (CSSS) model \citep{Zhao_1995_JGR}, magnetohydrostatic (MHS) model \citep{Bogdan_1986_ApJ, Neukirch_1995_AandA}, and nonlinear force-free field (NLFFF) model \citep{Wiegelmann_2004_SolPhys, Wiegelmann_2007_SoPh, Contopoulos_2011_SoPh}, and magneto-frictional equilibrium with solar wind outflow \citep{Rice_2021_ApJ}}. Among them, \rev{PFSS model, which assumes a current-free corona ($\nabla \times \boldsymbol{B} = 0$) and imposes a radial field condition at a spherical source surface,} is widely used and is adopted in various heriospheric models \citep{Usmanov_1993_SoPh, Riley_2001_JGR, Odstrcil_2003_AdSpR, Shiota_2014_SpaceWeather, Shiota_2016_SpaceWeather, Pomoell_2018_JSWSC}. Despite its simplicity and low computational cost, PFSS appears to \rev{be capable of reproducing} magnetic-field structures comparable to 3D MHD models \citep{Riley_2001_JGR, Reville_2020_ApJ, Huang_2024_ApJ} when the source-surface height, beyond which all field lines are \rev{assumed to be} open, is properly set.

While the source-surface height ($R_{\rm SS}$) is commonly set to $2.5R_\odot$, it is known to fluctuate with the phase of solar magnetic activity and should be adjusted accordingly \citep{Lee_2011_SoPh, Arden_2014_JGR, Nikolic_2019_SpWea, Benavitz_2024_ApJ, Huang_2024_ApJ}. A key issue is the absence of agreement on how to make this adjustment. Table~\ref{table:previous_works} lists studies that optimized $R_{\rm SS}$, along with their analysis periods, magnetograms, optimization metrics, and results. Regarding dependence on magnetic activity, some studies claim that $R_{\rm SS}$ is larger during solar minimum \citep{Arden_2014_JGR, Benavitz_2024_ApJ}, while others argue it is larger during solar maximum \citep{Nikolic_2019_SpWea, Huang_2024_ApJ}. Consequently, we have yet to determine how $R_{\rm SS}$ should be set or what physical quantity characterizes it. 

In this study, we analyze the dependence of $R_{\rm{SS}}$ on magnetic activity and aim to establish an empirical relation. To minimize the biases from magnetogram artifacts and limited analysis periods, we perform the same analysis over the longest possible time span using a wide variety of magnetograms. We then discuss the dependence of $R_{\rm{SS}}$ behavior on the magnetogram used and select the most appropriate one. Based on this selection, we derive an empirical relation to estimate $R_{\rm{SS}}$ from magnetogram data and approach the physical quantities characterizing $R_{\rm{SS}}$.

\section{Data analysis} \label{sec:data_analysis}

We apply the PFSS method to photospheric magnetic field maps with various source-surface radii ($R_{\rm SS}$) and estimate the optimal $R_{\rm SS}$ by comparing the resulting open flux with that derived from in-situ observations \citep{Lee_2011_SoPh, Arden_2014_JGR, Nikolic_2019_SpWea}. For simplicity, we assume a conventional spherical source surface. While multiple studies suggest that the source surface is non-spherical \citep{Schulz_1978_SoPh, Levine_1982_SoPh, Boe_2020_ApJ, Panasenco_2020_ApJS, Kruse_2021_AandA}, we do not consider such effects. Assuming a spherical source surface simplifies the discussion by making $R_{\rm SS}$ the only characteristic parameter in the PFSS method. 


The choice of magnetogram used as input in the PFSS method is crucial, as the PFSS solution is highly sensitive to the strength of the polar magnetic field \citep{Riley_2019_ApJ, Wang_2022_ApJ}. This field varies significantly depending on the resolution of the magnetogram, signal-to-noise ratio, and pole-filling techniques \citep{Svalgaard_1978_SoPh, Sun_2011_SoPh}. Since a ground truth for magnetograms is still unavailable \citep{Riley_2014_SoPh}, this study applies the PFSS method to various magnetograms and selects the most suitable one based on the results. Specifically, we use data from the Kitt Peak Vacuum Telescope \citep[KPVT,][]{Jones_1992_SolPhys}, Synoptic Optical Long-term Investigations of the Sun/Vector Spectromagnetograph \citep[SOLIS/VSM,][]{Keller_2003_ASPC, Henney_2009_ASPC}, the Global Oscillation Network Group \citep[GONG,][]{Hill_2018_SpWea}, and the Helioseismic and Magnetic Imager \citep[HMI,][]{Schou_2012_SoPh} onboard the Solar Dynamics Observatory \citep[SDO,][]{Pesnell_2012_SolPhys}. \rev{We also use their ADAPT realizations, which are global photospheric magnetic field maps constructed from the observations through a data-assimilation model} \citep{Worden_2000_SoPh, Arge_2010_AIPC, Arge_2013_AIPC, Hickmann_2015_SoPh}. In addition, we analyze data from the Michelson Doppler Imager \citep[MDI,][]{Scherrer_1995_SoPh} on the Solar and Heliospheric Observatory \citep[SOHO,][]{Domingo_1995_SoPh} spacecraft.

\begin{figure}
    \plotone{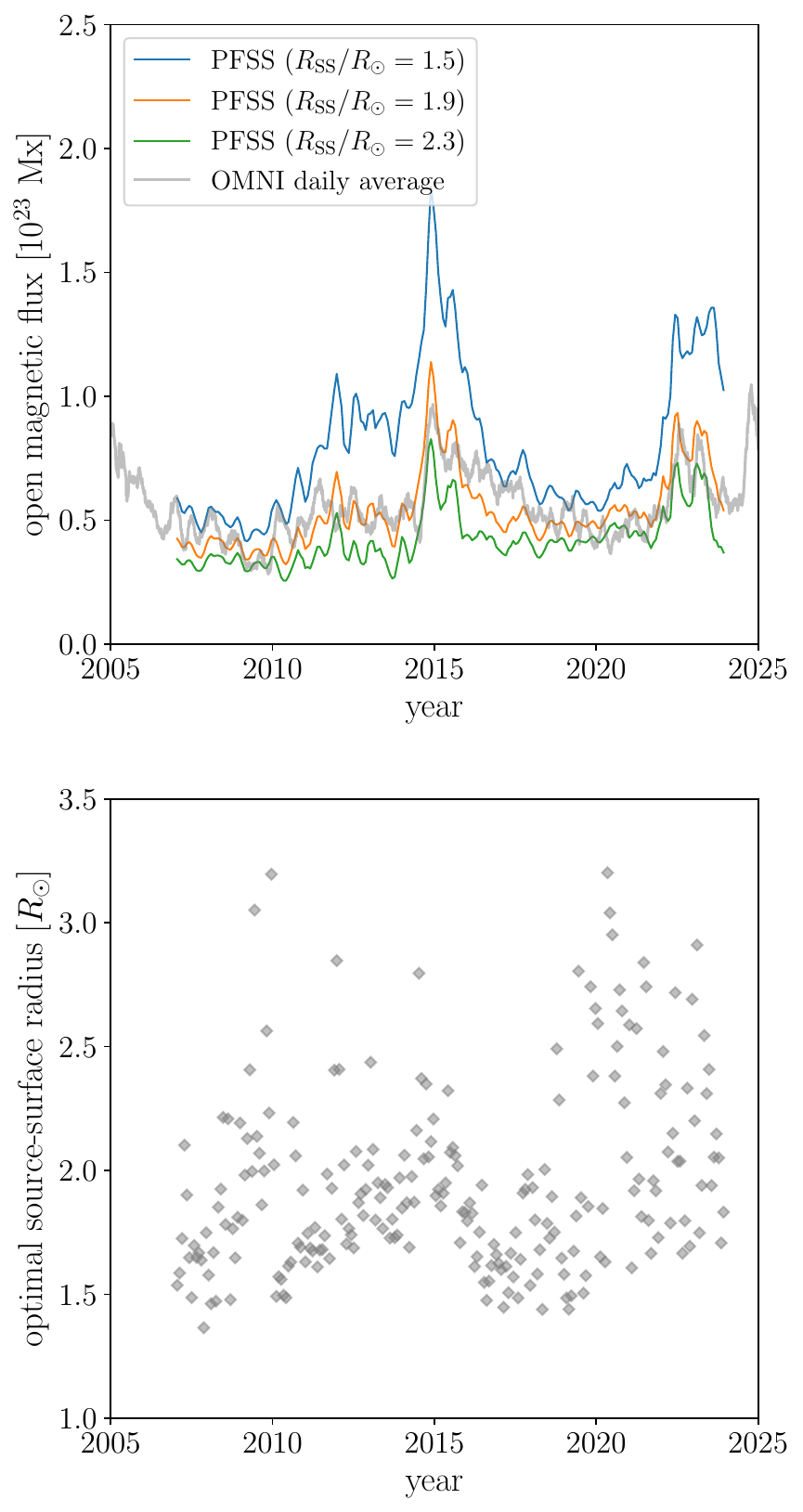}
    \caption{
    Process of calculating the optimal value of the source-surface height ($R_{\rm SS}^{\rm opt}$). The top panel presents the time evolution of the 3-rotation averaged open fluxes estimated from in-situ observations (gray) and the PFSS extrapolations using ADAPT-GONG as input, with different source-surface radii (blue: $R_{\rm SS}/R_\odot = 1.5$, orange: $R_{\rm SS}/R_\odot = 1.9$, green: $R_{\rm SS}/R_\odot = 2.3$). The bottom panel presents the temporal evolution of the optimal source-surface height (ADAPT-GONG) computed using Equation~\eqref{eq:rssopt_interpolation}.
    \label{fig:rssopt_calc_example_adapt-gong}}
\end{figure}

\begin{figure*}
    \plotone{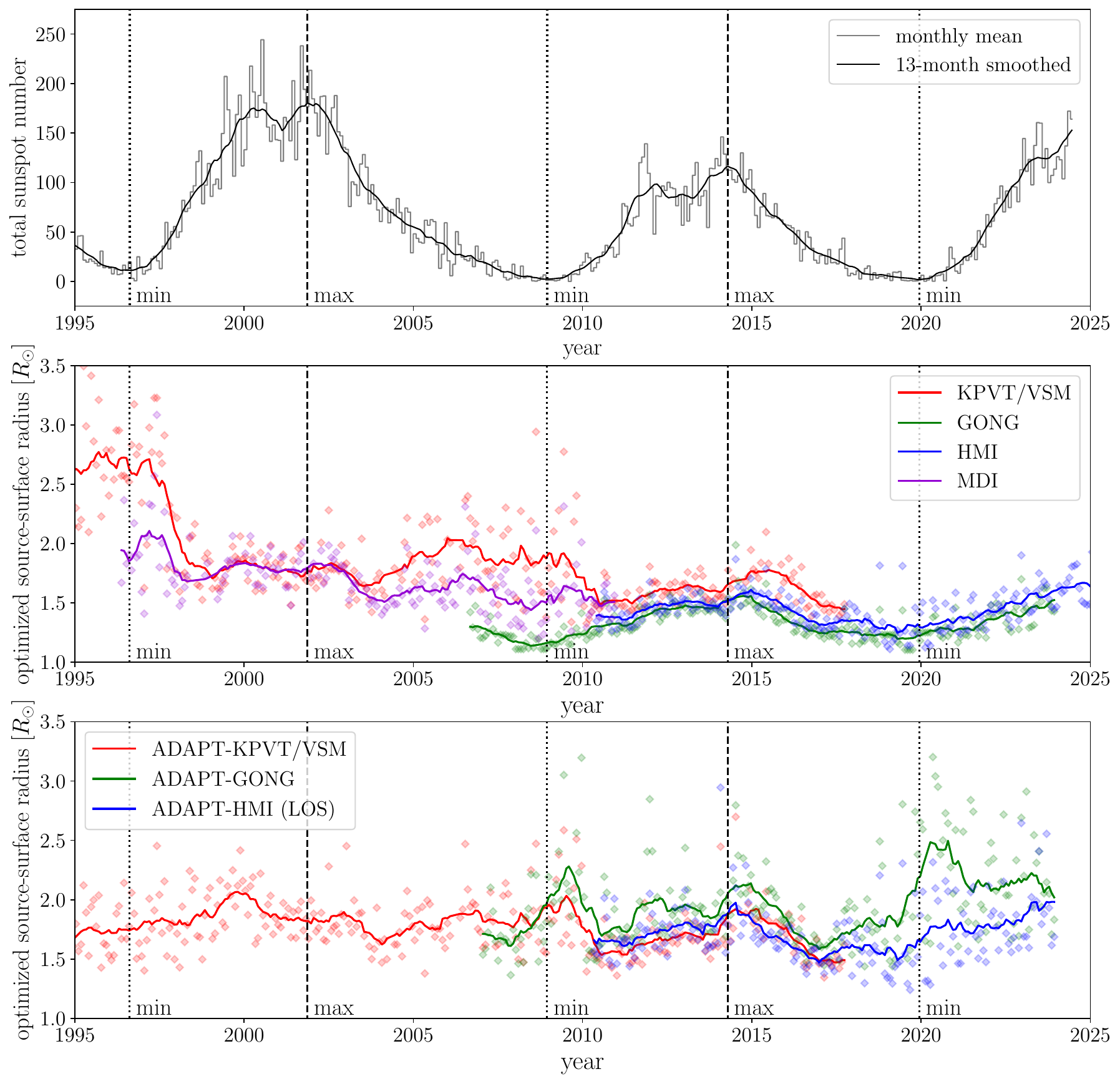}
    \caption{
    Long-term evolution of solar magnetic activity and optimal source-surface height derived from various magnetograms. Top: monthly-averaged (gray) and 13-month smoothed (black) sunspot numbers. Middle and bottom: optimal source-surface heights estimated from observed (middle) and ADAPT (bottom) magnetograms. Red, purple, green, and blue lines denote KPVT, SOHO/MDI, GONG, and SDO/HMI, respectively. Transparent diamonds show values for each Carrington Rotation; solid lines indicate 13-CR averages.
    \label{fig:evolution_of_rssopt_observed_and_adapt_maps}}
    \vspace{1em}
\end{figure*}

The optimal value of $R_{\rm SS}$ was determined using the following procedure. For each Carrington Rotation, we performed the PFSS extrapolation with the open-source software {\it pfsspy} \citep{Stansby_2020_JOSS}, using the synoptic map of the photospheric radial magnetic field as input. When utilizing the ADAPT map, we selected the synoptic map corresponding to the central date of the Carrington Rotation. ADAPT includes 12 realizations to account for uncertainties in surface convection parameters. Since solar wind parameters can sometimes significantly depend on the realization, ensemble simulations using all realizations are often performed \citep{da_Silva_2023_SpWea, Perri_2023_ApJ}. However, in the present study, the dependence is minor (see Appendix~\ref{app:realization_dependence}), and thus only the first realization is used. 

\rev{
We varied the source‐surface radius $R_{\rm SS}$ from 1.1 to 3.5 (in steps of 0.2).
For each case, the open flux (denoted as $\Phi_{\rm PFSS}$) was computed from the PFSS solution as
\begin{equation}
    \Phi_{\rm PFSS} = R_{\rm SS}^{2} \int_{4\pi} \left| B_r(R_{\rm SS},\theta,\phi) \right| \, {\rm d}\Omega, \label{eq:pfss_open_flux}
\end{equation}
where $B_r(R_{\rm SS},\theta,\phi)$ is the radial field on the source surface returned by {\it pfsspy}, and ${\rm d}\Omega = \sin\theta \, {\rm d}\theta \, {\rm d}\phi$. The colored lines in the top panel of Figure~\ref{fig:rssopt_calc_example_adapt-gong} represent examples of PFSS-derived open fluxes, averaged over three Carrington Rotations, obtained using the ADAPT-GONG synoptic map.
}

The observed open flux \rev{(denoted as $\Phi_{\rm H}$)} for the corresponding Carrington Rotation is determined using the following equation \citep{Badman_2021_AandA}:
\begin{align}
    \Phi_{\rm H} = \frac{1}{P_{\rm rot}} \int_{\tau_{\rm st}}^{\tau_{\rm ed}} {\rm d}t \ 4 \pi r_{\rm E}^2 \left| \overline{B_{r, {\rm E}}} (t) \right|, \label{eq:open_flux_observation}
\end{align}
where $P_{\rm rot}$ represents the solar rotation period, $\tau_{\rm st}$ and $\tau_{\rm ed}$ indicate the start and end times of the target Carrington Rotation, $r_{\rm E} = 1 {\rm \ au}$ is the heliocentric distance of the Earth, and $B_{r, {\rm E}}$ is the daily average value for the radial field strength near 1 au retrieved from OMNI database (\url{https://omniweb.gsfc.nasa.gov/}). When computing the integral in Equation~\eqref{eq:open_flux_observation}, we do not perform fractional-day adjustments at the endpoints. For example, for Carrington Rotation 2100 (from 14:45 on 9 August 2010 to 20:35 on 5 September 2010), the integral in Equation~\eqref{eq:open_flux_observation} is approximated as a simple arithmetic average over the 27 daily data points from 10 August to 5 September. We also neglect the time lag for solar wind propagation to 1 au. Although there is typically a 3–5 day delay for the variation in the coronal magnetic field to affect the field at 1 au, we assume this delay introduces a negligible error due to the averaging over each Carrington rotation. However, when discussing daily variations rather than rotation-averaged $R_{\mathrm{ss}}$, accounting for the solar wind travel time becomes essential, given the significant daily variability in the optimal source surface height (see Appendix~\ref{app:daily_variation}).

The open-flux estimation by Equation~\eqref{eq:open_flux_observation} is supported by the Ulysses observations, which show that $B_r r^2$ remains nearly constant with latitude \citep{Smith_1995_GRL, Smith_2003_AIPC}. In fact, the open flux estimated during the fast latitude scans of Ulysses agrees with that from Equation~\eqref{eq:open_flux_observation} within a $5 \%$ error \citep{Lockwood_2004_AnGeo}. We note, however, that $\left|B_r\right| r^2$ exhibits a slight radial dependence \citep{Owens_2008_JGRA, Badman_2021_AandA}, likely due to the folded field lines \citep{Lockwood_2009_JGRA_1, Lockwood_2009_JGRA_2} including switchbacks \citep{Bale_2019_Nature, Kasper_2019_Nature}, which yields the overestimation of the open flux \citep{Owens_2017_JGR, Frost_2022_SoPh}. Given that this flux excess effect is negligible when averaging over timescales of  approximately 20 hours \citep{Frost_2022_SoPh}, and since we use daily averages in estimating the open flux, we regard the error due to this effect as minimal. The grey line in the top panel of Figure~\ref{fig:rssopt_calc_example_adapt-gong} illustrates the evolution of the observed open flux, averaged over three Carrington Rotations.

For each Carrington Rotation, the optimal $R_{\rm SS}^{\rm opt}$ is calculated as follows. Let $\Phi_{\rm PFSS}^i$  be the modeled open flux corresponding to the $i$-th source-surface height (denoted as $R_{\rm SS}^i$). 
We note that a larger source-surface height leads to a smaller PFSS open flux, as is evident from the definition of the source surface. When an index $i$ is found such that $\Phi_{\rm PFSS}^i > \Phi_{\rm H}$ and $\Phi_{\rm PFSS}^{i+1} < \Phi_{\rm H}$ (where $\Phi_{\rm PFSS}^i$ decreases as $R_{\rm SS}^i$ increases), the optimal $R_{\rm SS}$ is obtained by a linear interpolation as follows:
\begin{align}   
    R_{\rm SS}^{\rm opt} = \frac{R_{\rm SS}^i \left( \Phi_{\rm H} - \Phi^{i+1}_{\rm PFSS} \right) + R_{\rm SS}^{i+1} \left( \Phi^i_{\rm PFSS} -  \Phi_{\rm H} \right)}{\Phi^{i}_{\rm PFSS} - \Phi^{i+1}_{\rm PFSS}}. \label{eq:rssopt_interpolation}
\end{align}
The bottom panel of Figure~\ref{fig:rssopt_calc_example_adapt-gong} shows the optimal source-surface height calculated using the above procedure, with the synoptic map from ADAPT-GONG.

\section{Result}

\begin{figure}
    \plotone{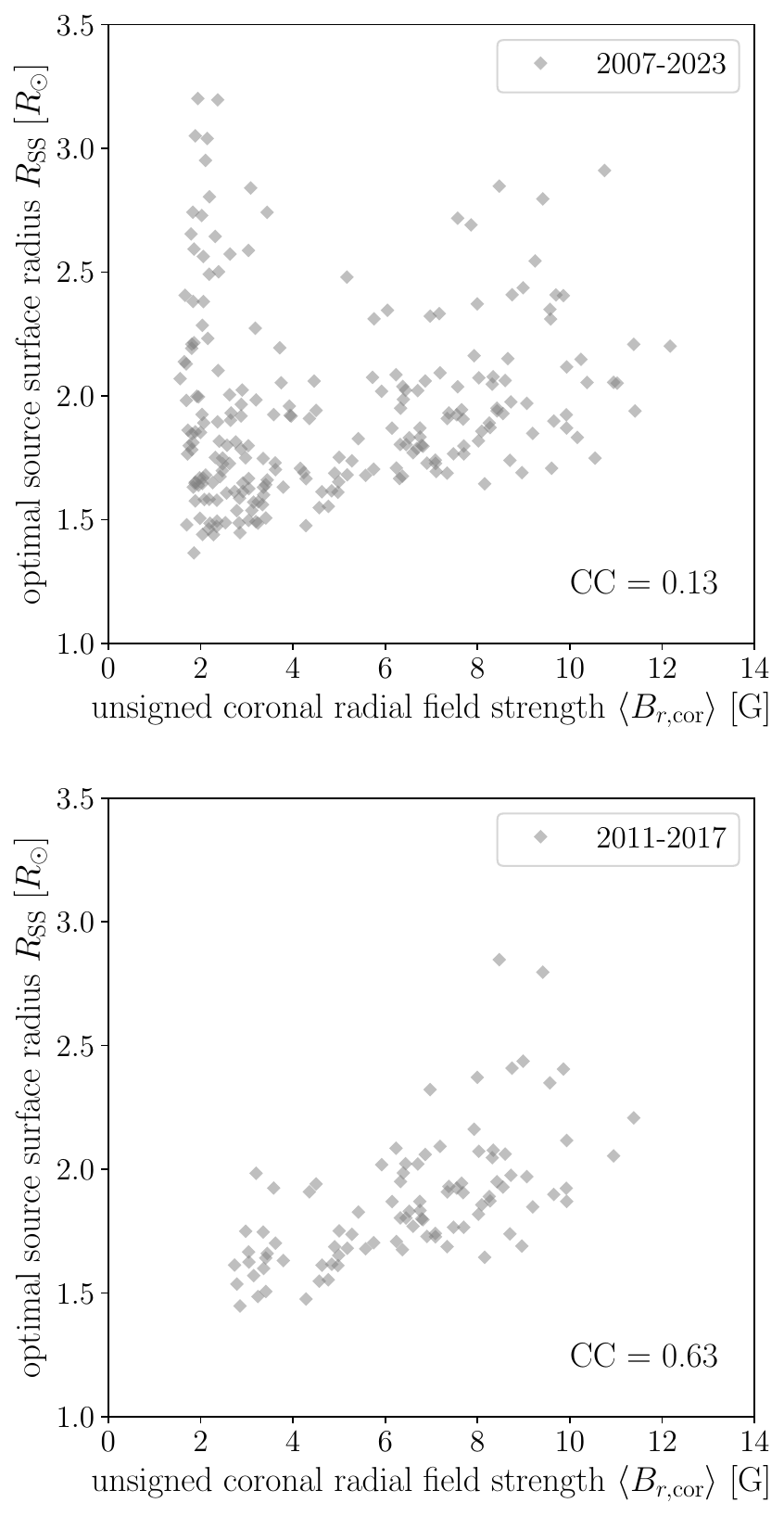}
    \caption{Relation between optimal source-surface height ($R_{\rm SS}^{\rm opt}$) and mean unsigned coronal radial field strength ($\left< B_{r, {\rm cor}} \right>$). Data from 2007 to 2023 are shown in the upper panel, and data from 2011 to 2017 in the lower panel. Each panel displays the value of the \rev{Peason correlation coefficient (PCC)}.
    \label{fig:comparison_with_Huang24_whole_and_limited_period}}
\end{figure}

\subsection{Optimal $R_{\rm SS}$ with various synoptic maps}

Figure~\ref{fig:evolution_of_rssopt_observed_and_adapt_maps} presents the long-term evolution of the sunspot number and the estimated optimal source-surface heights based on different synoptic maps. The top panel displays the sunspot number, obtained from the World Data Center SILSO, Royal Observatory of Belgium, Brussels \citep[][\url{https://doi.org/10.24414/qnza-ac80}]{SILSO_Sunspot_Number}. The middle and bottom panels show the results based on observed synoptic maps and ADAPT maps, respectively. Color variations indicate the differences in the input magnetograms. We note that the qualitatively different behavior of $R_{\rm SS}^{\rm opt}$ derived from KPVT/VSM and ADAPT-KPVT/VSM before 1998 is likely due to calibration issues in KPVT \citep{Wallace_2019_SoPh}, suggesting that the discrepancy is probably an artifact.

In the middle panel of Figure~\ref{fig:evolution_of_rssopt_observed_and_adapt_maps}, although all magnetograms exhibit similar behavior after 2011, a significant difference is observed before 2011, particularly near the solar activity minima. During solar minima, the large-scale field is dominated by the axial dipolar field, making the PFSS results highly sensitive to polar field observations \citep{Wang_2009_SSRev, Wang_2022_ApJ}. This sensitivity likely explains the discrepancies among the magnetograms, as the polar field strength depends on the telescope's spatial resolution \citep{Petrie_2022_ApJ, Sinjan_2024_AandA} and the method of polar-field interpolation \citep{Sun_2011_SoPh, Petrie_2015_LRSP}, both of which vary across magnetograms.


\rev{Conversely, as shown in the bottom panel of Figure~\ref{fig:evolution_of_rssopt_observed_and_adapt_maps}, the behavior of $R_{\rm SS}^{\rm opt}$ derived from ADAPT maps is more consistent across different input magnetograms than that derived from the observed synoptic maps. This is because the ADAPT model estimates the polar field based on physical flux-transport processes \citep{Arge_2010_AIPC, Hickmann_2015_SoPh}, rather than artificial interpolation, thereby providing a more homogeneous treatment of the polar regions. Considering both the reduced magnetogram dependence and the physically motivated polar-field estimation, the use of ADAPT maps is the most reasonable choice for our analysis.}

\rev{At the same time, we note that non-negligible differences can still arise among ADAPT products. In particular, around the solar cycle 24 minimum, the optimal source-surface height inferred from ADAPT-GONG and ADAPT-HMI can differ by about $1R_{\odot}$. This difference is likely caused by differences in the realization procedure of the ADAPT model \citep{Barnes_2023_ApJ} rather than discrepancies in the basal magnetograms themselves, as such a large offset is not observed when using the observed GONG and HMI synoptic maps as input for PFSS. As there is no ground truth for the photospheric magnetic field, we cannot determine which ADAPT product is closer to reality.}

\rev{For this study, we adopt ADAPT-GONG as the input magnetogram because it provides the longest and most continuously updated coverage—from 2006 to 2023 ($\sim$17 years, spanning two solar minima and one maximum)—allowing a statistically robust analysis over more than one solar cycle. Such long-term analysis would not be possible with ADAPT-HMI (available only since 2010) or ADAPT-KPVT/VSM (terminating in 2017).}

\subsection{Apparent dependence on the activity level: effect of analysis period}

As shown in the bottom panel of Figure~\ref{fig:evolution_of_rssopt_observed_and_adapt_maps}, the $R_{\rm SS}^{\rm opt}$ calculated using ADAPT-GONG increases during both solar maximum and minimum, showing no monotonic dependence on solar activity. In contrast, a recent study by \citet{Huang_2024_ApJ} suggests that $R_{\rm SS}$ increases monotonically with activity level—rising near solar maximum and decreasing near minimum. The primary reason for this difference may lie in the distinct approaches used to optimize $R_{\rm SS}$; while our study bases the optimization on in-situ observations, \citet{Huang_2024_ApJ} rely on the output of the AWSoM model \citep{van_der_Holst_2014_ApJ} (see Table~\ref{table:previous_works}). In addition to this methodological difference, the analysis periods also differ, which could influence the conclusions. We therefore examine how the choice of analysis period affects the behavior of $R_{\rm SS}$.

\citet{Huang_2024_ApJ} suggest that the source-surface height correlates with the mean radial field strength at the coronal base, defined as
\begin{align}
    \left< B_{r, {\rm cor}} \right> (t) = \frac{1}{4\pi} \int {\rm d} \Omega \left| B_r (r_{\rm cb}, L, \phi, t) \right|,
\end{align}
where $r_{\rm cb}$ is the radial distance of the coronal base, set to $r_{\rm cb}/R_\odot = 1.01$. \rev{Here, $B_r(r_{\rm cb},\theta,\phi,t)$ was obtained from the output of {\it pfsspy}}. $t$ denotes time, $L$ heliographic latitude, and $\phi$ Carrington longitude. Although their study used an integral averaged over the closed-field region, we verified that this value is nearly the same as the global average, and therefore adopt $\left< B_{r, {\rm cor}} \right>$ to compare our result with \citet{Huang_2024_ApJ}.

The top panel of Figure~\ref{fig:comparison_with_Huang24_whole_and_limited_period} presents a scatter plot illustrating the relationship between the optimal source-surface height and the averaged coronal radial field strength over the entire analysis period (2006–2023). As shown in the figure, while some data points exhibit a monotonic increase with $\left< B_{r, {\rm cor}} \right>$, a significant fraction around $\left< B_{r, {\rm cor}} \right> = 2 {\rm \ G}$ remains uncorrelated, displaying a vertically elongated distribution. These points correspond to the peak in $R_{\rm SS}^{\rm opt}$ near the solar minimum, as seen in the bottom panel of Figure~\ref{fig:evolution_of_rssopt_observed_and_adapt_maps}. \rev{In fact, the correlation between $R_{\rm SS}^{\rm opt}$ and $\left< B_{r, {\rm cor}} \right>$ is weak, with a Pearson correlation coefficient (PCC) of 0.13, where the PCC measures the strength of a linear relationship between two variables, with 1 indicating a perfect linear correlation and 0 indicating no linear correlation.}

As shown in the bottom panel of Figure~\ref{fig:comparison_with_Huang24_whole_and_limited_period}, on the other hand, a clear correlation between $R_{\rm SS}$ and $\left< B_{r, {\rm cor}} \right>$ emerges when focusing on the solar cycle 24 maximum (2011–2017) with the \rev{PCC} of 0.63. Since 7 of the 9 Carrington Rotations studied by \citet{Huang_2024_ApJ} fall within this shorter period, the discrepancy between our results may partly reflect the difference in the analysis interval.

Our findings here can be summarized as follows. Over the full analysis period, $R_{\rm SS}$ and $\left< B_{r, {\rm cor}} \right>$ show no significant correlation. However, a clear correlation appears when focusing on the solar maximum, in agreement with \citet{Huang_2024_ApJ}. This implies that the analysis period must be sufficiently longer than a solar cycle to properly characterize $R_{\rm SS}$. Accordingly, fitting $R_{\rm SS}$ with $\left< B_{r, {\rm cor}} \right>$ or equivalent parameters is appropriate only near solar maximum, and different indicators are required around solar minimum. We explore such parameters in the following section.

\begin{figure*}
    \plotone{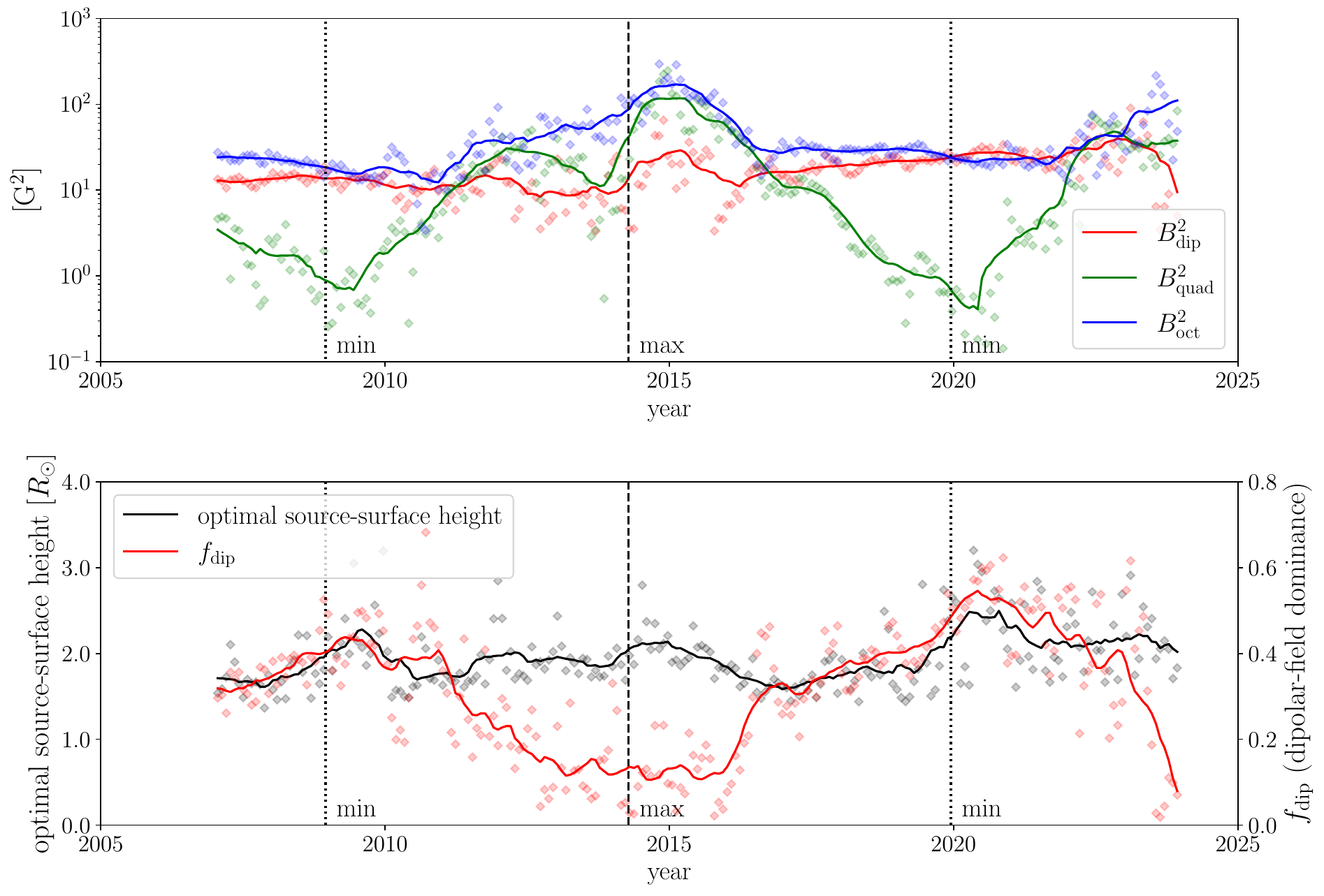}
    \caption{
    Time evolution of the multipolar components of the photospheric magnetic field and their relation to $R_{\rm SS}$. Top: Time evolution of dipolar (red, defined in Equation~\eqref{eq:dipolar_field_definition}), quadrupolar (green, Equation~\eqref{eq:quadrupolar_field_definition}), and octupolar (blue, Eq. (3)) field strengths. Bottom: Comparison of the time evolution of dipolar field dominance $f_{\mathrm{dip}}$ (red, Eq. (4)) and optimal source-surface height $R_{\mathrm{ss}}^{\mathrm{opt}}$ (black). In both panels, transparent diamonds show values for each Carrington rotation; solid lines represent the 13-CR running average.
    \label{fig:rss_fdip_long_term_comparison_adapt_gong}}
    \vspace{1em}
\end{figure*}

\begin{figure*}
    \plotone{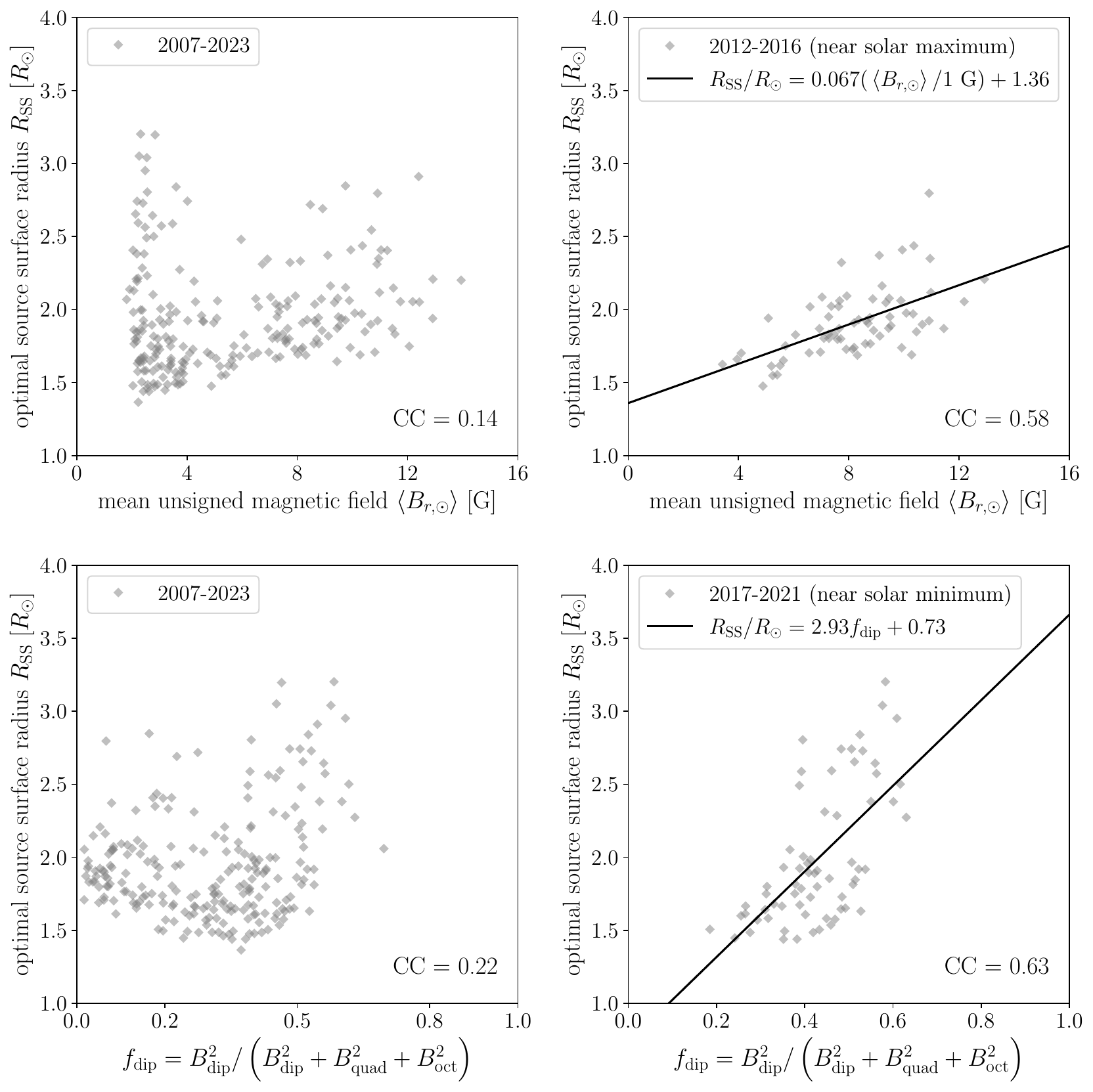}
    \caption{Scatter plots between the optimal source-surface height and characteristic parameters. The top two panels show $R_\mathrm{ss}^\mathrm{opt}$ versus the mean unsigned radial field strength at the photosphere, while the bottom panels show $R_\mathrm{ss}^\mathrm{opt}$ versus the dominance of the dipole component. The left column covers the full period (2007–2023), and the right column shows results for solar maximum (2012–2016) and minimum (2017–2021) phases.
    \label{fig:separate_scaling_adapt_gong}}
    \vspace{1em}
\end{figure*}

\begin{figure*}
    \plotone{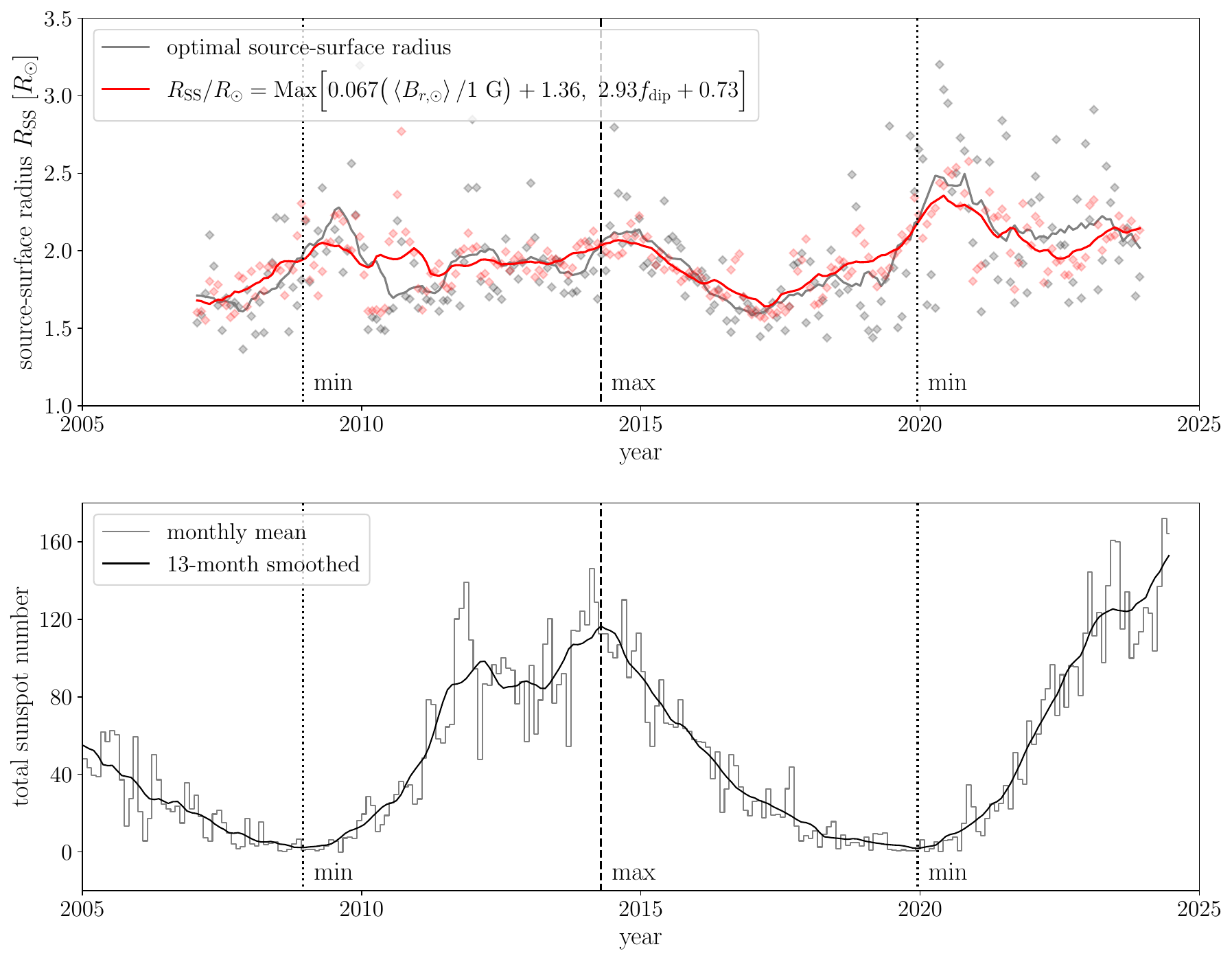}
    \caption{
    \rev{Long-term evolution of the optimal source-surface height and solar activity. Top panel:} comparison between the optimal source-surface height ($R_{\rm SS}^{\rm opt}$, black symbols and line) and the source-surface height given by Equation~\ref{eq:empirical_hybrid} (red symbols and line). Transparent diamonds show values for each Carrington Rotation; solid lines indicate 13-CR averages. \rev{Bottom panel: same as the top panel of Figure 2, but limited to the time range shown here.}
    \label{fig:evolution_of_rssopt_with_new_scaling_adapt_gong}}
    \vspace{1em}
\end{figure*}

\begin{figure}
    \plotone{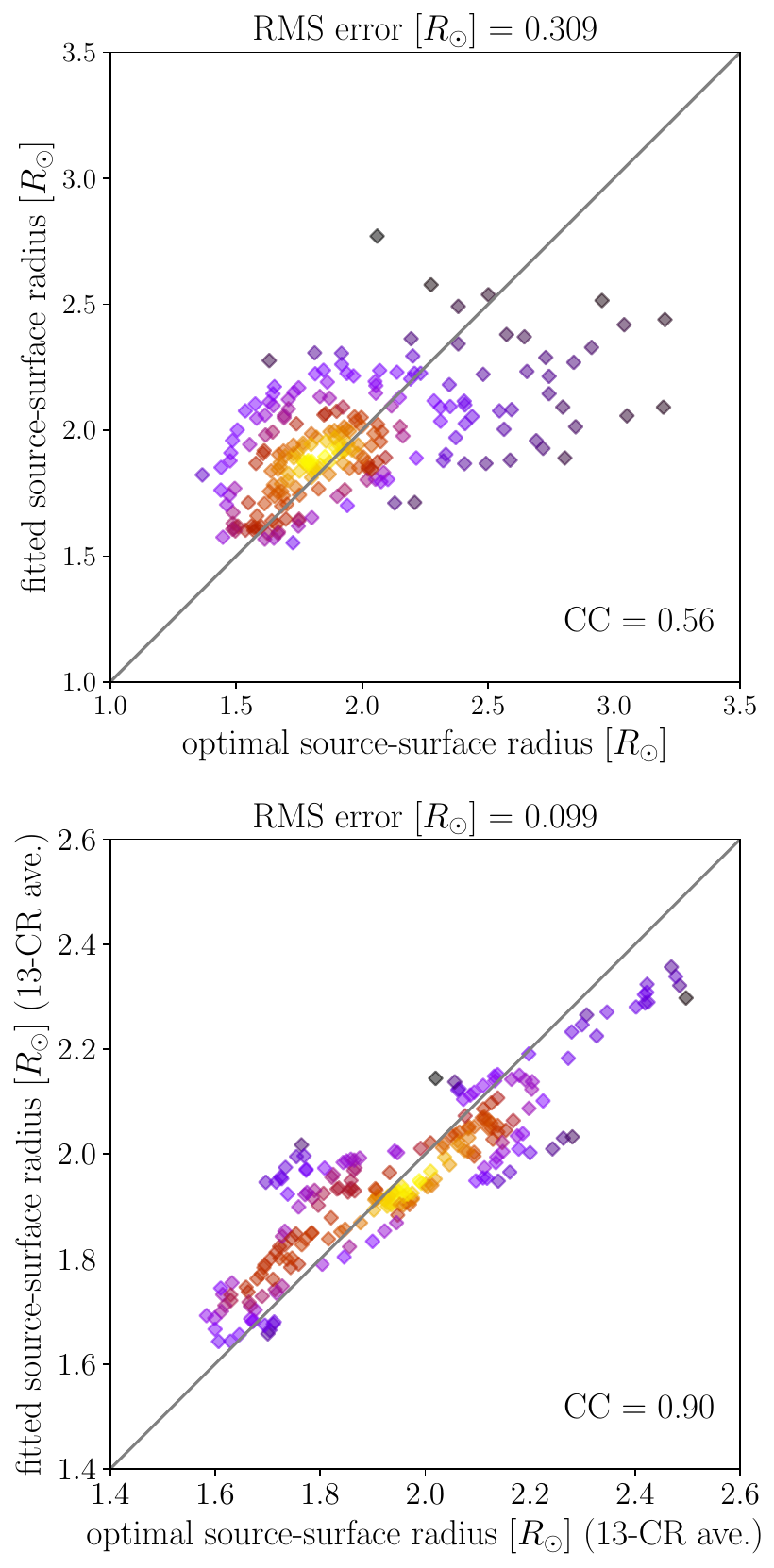}
    \caption{
    Scatter plot comparing the optimal source-surface height ($R_{\rm SS}^{\rm opt}$, black symbols and line) with the empirical source-surface height (Equation~\ref{eq:empirical_hybrid}) for each CR. The horizontal axis represents $R_{\rm SS}^{\rm opt}$, while the vertical axis shows the empirical estimate. The symbol color indicates data density, with brighter colors representing higher density. The top panel shows the direct comparison without averaging, whereas the bottom panel presents results after applying a 13-CR running average.
    \label{fig:rssopt_scatter_plot_optimal_vs_fitted_adapt_gong}}
    \vspace{1em}
\end{figure}

\subsection{Dependence on the large-scale field topology}

We focus on the large-scale magnetic field topology as a possible cause of the increased optimal source-surface height during solar minimum. The multipole components of the solar magnetic field vary with the activity cycle, with the (axial) dipolar field becoming dominant near solar minimum \citep{Virtanen_2017_AandA, Chu_2023_SoPh, Petrie_2024_JSWSC}. Since the dipole mode decays most slowly with distance from the Sun \citep{Petrie_2015_LRSP}, a larger dipole dominance (for a fixed photospheric field strength) allows the magnetic field to persist farther out. This suggests that a dipole-dominated configuration resists outward dragging by the solar wind more effectively, resulting in a higher source-surface height near solar minimum.

To investigate the relationship between $R_{\rm SS}$ and the large-scale magnetic field, particularly between dipolar-field dominance and $R_{\rm SS}$,, we calculate the dipolar, quadrupolar, and octupolar field strengths as follows \citep{Wang_2014_SSRev, Chu_2023_SoPh}.
\begin{equation}
    B_{\rm dip} = \sqrt{ \sum_{m=-1}^1 \left| \int {\rm d}\Omega \ B_{r,\odot} (\theta, \phi) Y_{1,m}^\ast \left( \theta, \phi \right) \right|^2}, \label{eq:dipolar_field_definition}
\end{equation}
\begin{equation}
    B_{\rm quad} = \sqrt{ \sum_{m=-2}^2 \left| \int {\rm d}\Omega \ B_{r,\odot} (\theta, \phi) Y_{2,m}^\ast \left( \theta, \phi \right) \right|^2}, \label{eq:quadrupolar_field_definition}
\end{equation}
\begin{equation}
    B_{\rm oct} = \sqrt{ \sum_{m=-3}^3 \left| \int {\rm d}\Omega \ B_{r,\odot} (\theta, \phi) Y_{3,m}^\ast \left( \theta, \phi \right) \right|^2}, \label{eq:octupolar_field_definition} 
\end{equation}
where $B_{r,\odot}$ denotes the radial field distribution on the photosphere, and $Y_{l,m}(\theta, \phi)$ represents the spherical harmonic function defined below.
\begin{align}
    Y_{l,m} \left( \theta, \phi \right) = \sqrt{\frac{2l+1}{4 \pi} \frac{(l-m)!}{(l+m)!}} e^{im\phi} P_{l,m} \left( \cos \theta \right).
\end{align}
We note that the definition of the associated Legendre polynomial $P_{l,m} (x)$ for a negative integer $m$ is given by
\begin{align}
    P_{l,m} (x) = (-1)^{m} \frac{(l+m)!}{(l-m)!} P_{l, -m} (x).
\end{align}
The above calculation is carried out using the open-source Python library {\it scipy} \citep{Virtanen_2020_NatMe}.

Figure~\ref{fig:rss_fdip_long_term_comparison_adapt_gong} shows the temporal evolution of the photospheric magnetic field (from ADAPT-GONG) multipolar components and their relation to $R_{\rm SS}^{\rm opt}$. The top panel presents the time variation of the multipolar components as defined in Equations~\eqref{eq:dipolar_field_definition}-\eqref{eq:octupolar_field_definition}. The dipolar field remains nearly constant regardless of the activity level, as the axial dipole strengthens during solar minimum  while the equatorial dipole intensifies during solar maximum \citep{Wang_2022_ApJ, Yoshida_2023_ApJ}, balancing each other. In contrast, higher-order components, in particular the quadrupolar field strength, vary significantly with activity level \citep{DeRosa_2011_IAUS}. Consequently, the relative dominance of the dipole field also changes accordingly.

To quantify the dominance of the dipolar field, we introduce the following parameter:
\begin{equation}
    f_{\rm dip} = \frac{B_{\rm dip}^2}{B_{\rm dip}^2 + B_{\rm quad}^2 + B_{\rm oct}^2}. \label{eq:dipolar_field_dominance}
\end{equation}
Physically, $f_{\rm dip}$ represents how dominant the dipole field is, in terms of energy, compared to the quadrupolar and octupolar fields. The bottom panel of Figure~\ref{fig:rss_fdip_long_term_comparison_adapt_gong} shows the temporal evolution of $f_{\rm dip}$ alongside that of $R_{\rm SS}^{\rm opt}$. Their comparison reveals that $f_{\rm dip}$ closely follows $R_{\rm SS}^{\rm opt}$ near solar activity minima, suggesting that $f_{\rm dip}$ is a suitable parameter to describe the behavior of $R_{\rm SS}$ during such periods.

\subsection{A new empirical formulation}

Based on the preceding analysis and its implications, we develop a new empirical relation for $R_{\rm SS}^{\rm opt}$ by independently deriving relations near the solar maximum and minimum. Near the solar maximum, $R_{\rm SS}^{\rm opt}$ tends to increase monotonically with activity, so we fit $R_{\rm SS}^{\rm opt}$ with the surface-averaged unsigned radial magnetic field defined below. 
\begin{equation}
    \left< B_{r,\odot} \right> = \int d\Omega \left| B_{r,\odot} \left( \theta, \phi \right) \right|.
\end{equation}
In contrast, near the solar minimum, we fit $R_{\rm SS}^{\rm opt}$ with $f_{\rm dip}$ (as defined in Equation~\eqref{eq:dipolar_field_dominance}). 

The top panels of Figure~\ref{fig:separate_scaling_adapt_gong} show the relation between $\left< B_{r,\odot} \right>$ and $R_{\rm SS}^{\rm opt}$ over the full analysis period (2007–2023, left) and near the solar maximum (2012–2016, right). No significant correlation is found over the entire period (\rev{PCC}: 0.14), whereas a significant positive correlation (\rev{PCC}: 0.58) appears near the solar maximum, indicating that $\left< B_{r,\odot} \right>$ can characterize $R_{\rm SS}^{\rm opt}$. In particular, near the solar maximum, 
\rev{the following relation is obtained from a least-squares fit:}
\begin{align}
     \frac{R_{\rm SS}^{\rm opt}}{R_\odot} = 0.067 \left( \frac{\left< B_{r,\odot} \right>}{1 {\rm \ G}} \right) + 1.36. \label{eq:empirical_solarmax}
\end{align}

The bottom panels of Figure~\ref{fig:separate_scaling_adapt_gong} display the relation between $f_{\rm dip}$ and $R_{\rm SS}^{\rm opt}$ for the full analysis period (2007–2023, left) and near solar minimum (2017–2021, right). We note that, while ADAPT-GONG contains data near the minimum of solar cycle 23, these data were excluded from the regression near solar minimum, as their reliability is reduced due to a strong dependence on the realization label (see Appendix A for detail). No significant correlation between $f_{\rm dip}$ and $R_{\rm SS}^{\rm opt}$ is found over the full period (\rev{PCC}: 0.22). In contrast, a significant positive correlation (\rev{PCC}: 0.63) is observed near solar minimum, suggesting that $f_{\rm dip}$ can serve as a predictive feature for $R_{\rm SS}^{\rm opt}$. Specifically, \rev{the following relation is obtained from a least-squares fit:}
\begin{align}
     \frac{R_{\rm SS}^{\rm opt}}{R_\odot} = 2.93 f_{\rm dip} + 0.63. \label{eq:empirical_solarmin}
\end{align}

Considering that $R_{\rm SS}^{\rm opt}$ is described by different empirical laws near solar maximum and minimum, and that each law underestimates $R_{\rm SS}^{\rm opt}$ when applied to the opposite phase, we propose a new empirical relation for the source-surface height by adopting the larger value of Equations~\eqref{eq:empirical_solarmax} and~\eqref{eq:empirical_solarmin} as follows:
\begin{align}
    \frac{R_{\rm SS}^{\rm opt}}{R_\odot} = {\rm Max} \bigg[ & 0.067 \left( \frac{\left< B_{r,\odot} \right>}{1 {\rm \ G}} \right) + 1.36, \ 2.93 f_{\rm dip} + 0.63 \bigg]. \label{eq:empirical_hybrid}
\end{align}
\rev{The top panel of} Figure~\ref{fig:evolution_of_rssopt_with_new_scaling_adapt_gong} presents the time evolution of $R_{\rm SS}^{\rm opt}$ predicted by Equation~\eqref{eq:empirical_hybrid} (red symbols and line) and $R_{\rm SS}^{\rm opt}$ inferred from observation (black symbols and line). \rev{To facilitate comparison with solar activity variation, the bottom panel displays the time evolution of the sunspot number.} The combination of empirical laws for the solar maximum and minimum allows us to reproduce the increasing trend of $R_{\rm SS}^{\rm opt}$ in both phases, confirming that the new scaling law is independent of solar activity.

To quantitatively assess the accuracy of our proposed empirical law, we present a scatter plot in Figure~\ref{fig:rssopt_scatter_plot_optimal_vs_fitted_adapt_gong} comparing the observation based $R_{\rm SS}^{\rm opt}$ with its empirical estimation (Equation~\eqref{eq:empirical_hybrid}). The top panel shows a one-to-one comparison for each \rev{Carrington Rotation (CR)}, while the bottom panel illustrates a 13-CR moving average. The color of the points represents the data density (kernel density estimate using Gaussian kernels), with brighter colors indicating higher density. As shown in the top panel, while our empirical law fails to reproduce some outlier cases with large $R_{\rm SS}^{\rm opt}$, it accurately captures the bulk population. The \rev{PCC} between the two is 0.56, and the root-mean-squared error is 0.31$R_\odot$. As shown in the bottom panel, the estimation accuracy improves with the 13-CR average, yielding a \rev{PCC} of 0.90 and a root-mean-squared error of 0.11$R_\odot$. Thus, our empirical relation reliably reproduces source-surface behavior on at least a yearly scale and typically predicts the source-surface for each CR with an uncertainty of approximately $\pm 0.3R_\odot$.

\section{Summary and Discussion}

In this study, we optimize the PFSS source-surface height based on the open flux estimated from the in-situ observations at 1 au and propose its empirical formulation. The key findings of this study are summarized as follows.
\begin{enumerate}
    \item As suggested in Figure~\ref{fig:evolution_of_rssopt_observed_and_adapt_maps}, long-term behavior of the optimal source-surface height significantly depends on the photospheric magnetic field used as input for PFSS extrapolation. In particular, when a synoptic map obtained directly from observations is used, the behavior of $R_{\rm SS}^{\rm opt}$ shows significant differences near the solar minimum. This is likely due to the variations in the pole-filling technique used in constructing the synoptic map. Indeed, in ADAPT maps, where similar pole-filling is applied, the behavior of $R_{\rm SS}^{\rm opt}$ remains consistent regardless of the basal magnetogram.
    \item Using ADAPT-GONG synoptic maps as input, we find that $R_{\rm SS}^{\rm opt}$ increases near both the activity maximum and minimum. Thus, the average unsigned radial field strength at the coronal base ($\left< B_{r, {\rm cor}} \right>$) alone cannot characterize $R_{\rm SS}^{\rm opt}$, as indicated by the top panel of Figure~\ref{fig:comparison_with_Huang24_whole_and_limited_period}. However, when the epoch is limited to the vicinity of solar maximum, a positive correlation between $\left< B_{r, {\rm cor}} \right>$ and $R_{\rm SS}^{\rm opt}$ is observed (bottom panel of Figure~\ref{fig:comparison_with_Huang24_whole_and_limited_period}), aligning with a previous study \citep{Huang_2024_ApJ}.
    \item The dipolar-field dominance ($f_{\rm dip}$, defined by Equation~\eqref{eq:dipolar_field_dominance}) is identified as another candidate parameter characterizing $R_{\rm SS}^{\rm opt}$ near the solar activity minimum, as shown in Figure~\ref{fig:rss_fdip_long_term_comparison_adapt_gong}. By fitting $R_{\rm SS}^{\rm opt}$ with $\left< B_{r, \odot} \right>$ near the activity maximum and with $f_{\rm dip}$ near the activity minimum, and then combining both, we derive an empirical relation (Equation~\eqref{eq:empirical_hybrid}) that reproduces the long-term trend of $R_{\rm SS}^{\rm opt}$, as shown in Figures~\ref{fig:evolution_of_rssopt_with_new_scaling_adapt_gong} and~\ref{fig:rssopt_scatter_plot_optimal_vs_fitted_adapt_gong}.
\end{enumerate}

The physical origin of our empirical formulation can be interpreted as follows. $R_{\rm SS}$ represents the location where magnetic field lines open, corresponding to the balance between the inertial force of the solar wind that tends to straighten the field lines and the magnetic tension that bends them back. Thus, if the magnetic field strength is maintained farther from the Sun, the closed magnetic field structure can resist the inertial force of the solar wind more effectively, leading to an increase in $R_{\rm SS}$. Assuming the spatial structure of the solar magnetic field (spectrum in terms of the spherical harmonic degree $l$) remains unchanged, a larger $\left< B_{r, \odot} \right>$ results in a stronger interplanetary magnetic field and a larger $R_{\rm SS}$. Conversely, if the surface magnetic field strength remains nearly constant, a more dominant dipolar field (a larger $f_{\rm dip}$) leads to a stronger field strength in the middle corona and a larger $R_{\rm SS}$. 

\rev{A key limitation of this study is that $R_{\rm SS}$ was optimized such that the open flux inferred from in-situ observations matches that obtained from the PFSS model. While this criterion is physically motivated, it is not unique. Optimizing $R_{\rm SS}$ based on other metrics, such as aligning coronal holes with open-field regions, is known to yield systematically larger values \citep[the open-flux problem;][]{Linker_2017_ApJ, Wang_2022_ApJ, Arge_2024_ApJ, Asvestari_2024_ApJ}. It should also be noted that recent studies suggest that open-field regions do not necessarily correspond to coronal holes \citep{Iijima_2025_arXiv}, casting doubt on the validity of optimizing $R_{\rm SS}$ based solely on coronal hole area. Consequently, our empirical relation may deviate from the true source-surface height and should be applied with caution.}

\rev{In addition, recent studies have reported mutually inconsistent solar-cycle trends in $R_{\rm SS}$, with some finding an increase near solar maximum \citep{Huang_2024_ApJ} and others near solar minimum \citep{Benavitz_2024_ApJ}. As discussed in Section~3.2, the discrepancy with \citet{Huang_2024_ApJ} may partly reflect their relatively short analysis period, which can bias the inferred long-term trend, whereas the difference from \citet{Benavitz_2024_ApJ} may stem from their use of only ten Carrington rotations, limiting the statistical robustness of their results. More fundamentally, the PFSS framework cannot determine which optimization criterion is physically most appropriate, and different choices naturally lead to differing conclusions \citep{Badman_2022_ApJ}.}

Despite the limitations discussed above, our results suggest that it is possible to optimize the source-surface height using only (current and past) magnetogram data, regardless of magnetic activity. This represents a significant step toward improving future space weather prediction models. Our empirical formulation is specific to ADAPT-GONG and must be carefully revised when using other magnetograms or data assimilation models. At the very least, this study indicates the need to consider two factors: the solar activity level and the power spectrum in terms of spherical harmonics.

\begin{acknowledgments}
This work was supported by JSPS KAKENHI Grant Numbers JP24K00688, JP25K00976, JP25K01052, and by the grant of Joint Research by the National Institutes of Natural Sciences (NINS) (NINS program No OML032402). Kyogo Tokoro is supported by International Graduate Program for Excellence in Earth-Space Science (IGPEES), a World-leading Innovative Graduate Study (WINGS) Program, the University of Tokyo.
\end{acknowledgments}

\begin{contribution}
MS was primarily responsible for data acquisition, analysis, and manuscript writing. KT developed the analysis programs, proposed data analysis methods, and provided detailed suggestions for improvement. DS and SI reviewed the entire manuscript and discussed the validity of the analysis methods.
\end{contribution}

\facilities{KPVT, SOLIS(VSM), SOHO(MDI), SDO(HMI)}

\software{matplotlib \citep{Hunter_2007_CSE}, NumPy \citep{Harris_2020_Nature}, scipy \citep{Virtanen_2020_NatMe}, Astropy \citep{AstropyCollaboration_2013_AandA, AstropyCollaboration_2018_AJ, AstropyCollaboration_2022_ApJ}, SunPy \citep{Mumford_2020_JOSS}, pfsspy \citep{Stansby_2020_JOSS}}

\appendix

\section{Dependence on ADAPT realization \label{app:realization_dependence}}

\begin{figure*}
    \plotone{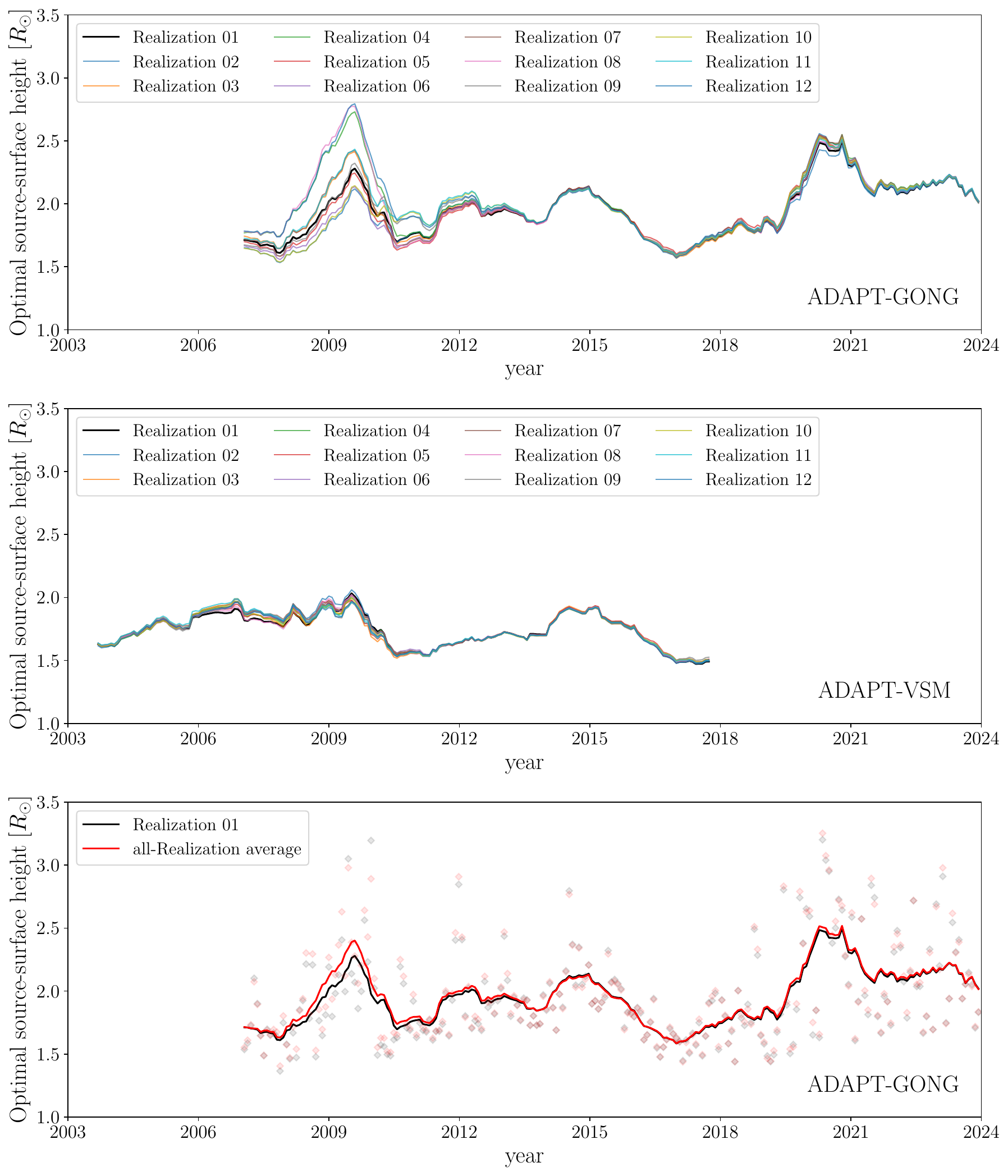}
    \caption{Upper panel: Dependence of the optimal source-surface height (13-CR average) on ADAPT realization, with different colors indicating different realizations. Lower panel: Comparison of the optimal source-surface height for Realization 01 used in this study (black) and the average over all realizations (red). Transparent diamonds show values for each CR; solid lines represent the 13-CR average.
    \label{fig:rssopt_realization_dependence}}
    \vspace{1em}
\end{figure*}

In this study, we consistently used the first realization of ADAPT (denoted hereafter as Realization 01). Since ADAPT includes 12 realizations based on different surface convection parameters, these differences could qualitatively impact magnetic field extrapolations. In this Appendix, we examine how much uncertainty the choice of ADAPT realization introduces, in particular in the derived optimal source-surface height.

The top panel of Figure~\ref{fig:rssopt_realization_dependence} shows the time evolution of the source-surface heights derived from different ADAPT-GONG realizations. For clarity, only the 13-CR running average is shown. The optimal source-surface height exhibits strong realization dependence around the cycle 23 minimum (in 2008). Although the GONG project began in 1995, its instruments were upgraded in 2006 \citep{Hill_2018_SpWea}, and ADAPT data assimilation has only been applied to observations since then. 
\rev{The polar fields observed at the 2008 solar minimum were largely produced by flux from bipolar regions that had emerged prior to 2006, consistent with the 2–5 year poleward transport timescale of polar field precursors \citep{Sheeley_2005_LRSP, Munoz-Jaramillo_2013_ApJ, Jiang_2014_SSRev}. This delayed transport likely accounts for the uncertainty in the ADAPT polar field realizations and in the optimal source surface heights derived from them.} \rev{In Figure~\ref{fig:rssopt_realization_dependence}}, from around 2013, the differences among realizations vanish, and a unique value is obtained for the optimal height during the cycle 24 minimum.

To assess the validity of using Realization 01 for analysis, the bottom panel of Figure~\ref{fig:rssopt_realization_dependence} compares the time evolution of the optimal source-surface height derived from Realization 01 with the mean across all realizations. The two agree well overall, including during the cycle 23 minimum where realization dependence is significant, suggesting that Realization 01 represents typical behavior. Notably, around the 2015 solar maximum, values from individual CRs (indicated by transparent diamonds) also align closely.

The implications drawn from the analysis of realization dependence are as follows. First, while the realization dependence near the solar cycle 23 minimum (using ADAPT-GONG) is relatively significant, it becomes negligible from the mid-2010s onward, at least in terms of determining the optimal source-surface height. Second, the mean optimal source-surface height across all realizations closely resembles that of Realization 01, indicating that Realization 01 represents a typical case among them. Therefore, the analysis based on Realization 01 is considered valid, although caution is warranted when interpreting data near the cycle 23 minimum.

\section{Daily variation in the optimal source-surface height \label{app:daily_variation}}

\begin{figure}[!t]
    \plotone{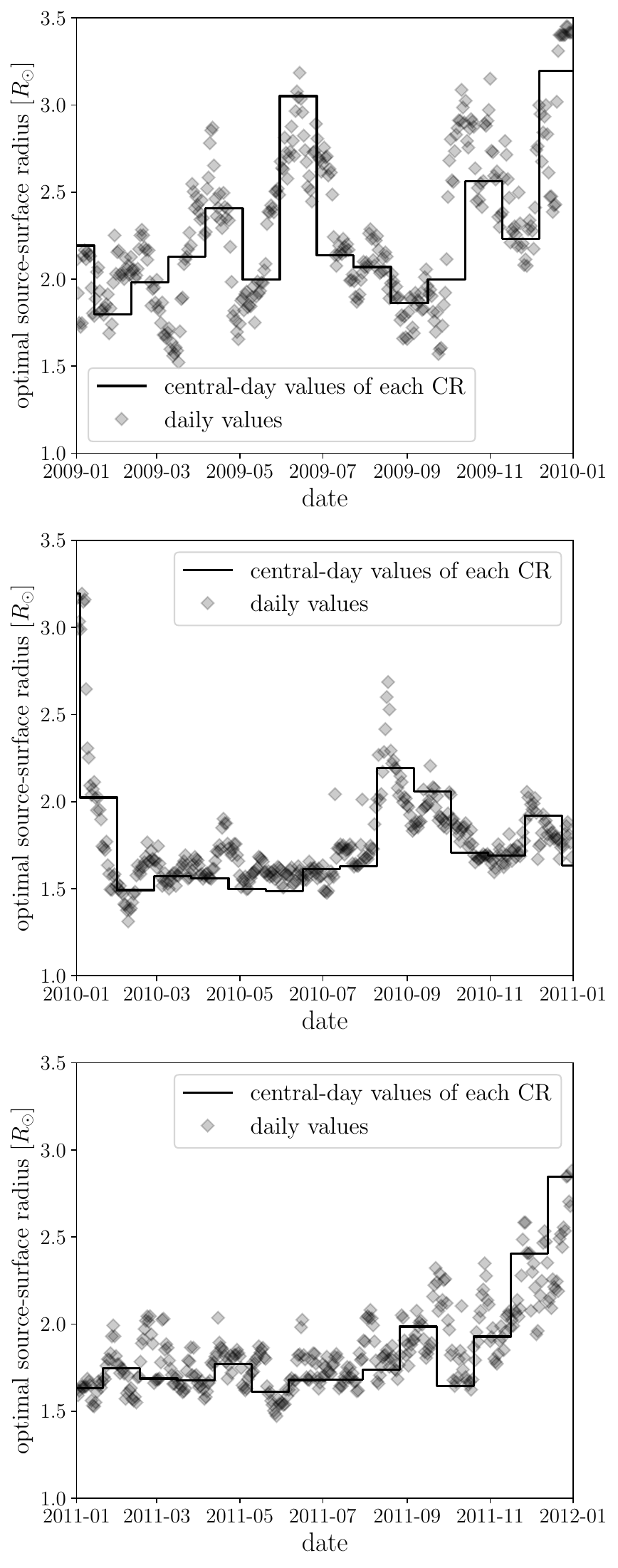}
    \caption{Daily variation of the optimal source-surface height calculated with ADAPT-GONG (Realization 01). The three panels show data for 2009, 2010, and 2011 from top to bottom. Transparent diamonds indicate the optimal source-surface height for each date, while the stepped solid line shows the value computed at the central date of each Carrington Rotation. By definition, the diamonds and the solid line coincide on the central dates.
    \label{fig:daily_analysis_of_rssopt_adapt_gong}}
\end{figure}

In calculating the optimal source-surface height for each Carrington rotation with the ADAPT maps, we used the data corresponding to the central date of the rotation. However, given the significant daily variation in the solar surface magnetic field and solar wind flux, this approach may be inadequate. In this Appendix, we quantitatively assess this variation by computing the optimal source-surface height for each day and evaluating the associated uncertainty.

Specifically, we estimate the optimal source-surface height for each day as follows. For each daily ADAPT-GONG map, we perform PFSS calculations using various source-surface heights and compute the corresponding open flux \rev{via Equation~\eqref{eq:pfss_open_flux}}. 
\rev{These PFSS-derived open flux values are then compared with the observed solar-wind open flux, which is calculated as a running average over one Carrington Rotation centered on the target date. The optimal $R_{\rm SS}$ is then calculated with Equation~\eqref{eq:rssopt_interpolation}.} We note that the solar wind open flux varies from day to day within the same Carrington rotation, and the value obtained on the central day matches that used in the main analysis.

Figure~\ref{fig:daily_analysis_of_rssopt_adapt_gong} compares the time evolution of the optimal source-surface height calculated from daily data (shown as diamonds) with that from the central date of each Carrington rotation (solid line). The daily variation of the optimal source-surface height is generally non-negligible; for instance, it changes by up to 1.5$R_\odot$ within a single rotation around April 2009 and January 2010. \rev{However, the significant day-to-day variation of the optimal source-surface height within a single Carrington rotation does not necessarily represent true physical fluctuations. This is because the large-scale photospheric magnetic field is expected to evolve only slowly over a few days, and such timescales are much shorter than those predicted by standard surface-flux-transport models \citep{Mackay_2012_LRSP, Jiang_2014_SSRev, Yeates_2023_SSRev}. Therefore, the observed variability likely reflects the intrinsic uncertainty of our approach. We quantify this uncertainty as a root-mean-squared daily variation of about 0.3$R_\odot$, which we regard as the precision limit of our method.}

\end{document}